\documentstyle[prl,aps,floats,twocolumn]{revtex}
\input psfig.tex
\begin{document}
\draft
\twocolumn[\hsize\textwidth\columnwidth\hsize\csname @twocolumnfalse\endcsname
\preprint{PUPT 94-1514}
\title{Temperature-Polarization Correlations from Tensor Fluctuations}
\author{ R.G. Crittenden$^1$, D. Coulson$^2$  \and N.G. Turok$^{1,3}$}
\address{$^{(1)}$ Joseph Henry Laboratory, Princeton University,
Princeton NJ, 08544\\
$^{(2)}$ Department of Physics, University of Pennsylvania, 
Philadelphia, PA 19104\\
$^{(3)}$ Isaac Newton Institute, University of Cambridge, 
Cambridge CB3 0EH }
\date{11/28/94}
\maketitle

\begin{abstract}
We study the polarization-temperature
correlations on the cosmic microwave sky 
resulting from an initial scale invariant 
spectrum of tensor (gravity wave) fluctuations, 
such as those which might arise during inflation.
The correlation function has the 
opposite sign to that for scalar fluctuations on large scales, 
raising the possibility of a direct determination 
of whether the microwave anisotropies have a significant
tensor component. We briefly discuss the important problem
of estimating the expected foreground contamination.
\\
\end{abstract}
]

COBE's detection 
of microwave background temperature
anisotropies~\cite{smoot92}  
has excited great interest in theories of cosmic structure formation.
Theoretical mechanisms for 
producing  CMB anisotropies  include
primordial energy density (scalar) 
and gravitational wave (tensor) fluctuations generated 
during inflation, or the gravitational
fields induced by 
cosmic defects.
These three  mechanisms
make surprisingly similar predictions for anisotropies 
on the large angular scales probed by COBE,
and  further observations are needed in order to discriminate
between them.
It is clearly of particular value to
identify  distinctive 
signals associated with 
specific physical effects, in order to determine in 
as direct a manner as possible which
mechanisms produced these anisotropies.

The level of temperature fluctuations on smaller scales 
provides the simplest such test.
For example, within inflationary models, 
the anisotropies
due to adiabatic scalar fluctuations generally 
increase on scales of order the horizon at last scattering 
(the ``Doppler peak") whereas those due to tensor
fluctuations decay away
~\cite{critt93}. 
However the height of the Doppler
peak also depends sensitively on several poorly determined
cosmological parameters (for example
the Hubble constant and the baryon density), 
on the ionization history of the universe, and on theoretical 
model parameters (the details of the inflaton potential~\cite{infl,edetal94}).
These uncertainties make it difficult
to unambiguously determine the size of the tensor
contribution~\cite{critt94}.

The polarization of the microwave sky could provide
invaluable extra information, not far beyond the reach of
current experiments. The degree of linear 
polarization expected from
scalar and tensor fluctuations produced during inflation
has been calculated
~\cite{k83,be,Polnarev,Ng93,cds}, with the result that for 
a given level of temperature anisotropy tensor perturbations
do produce a somewhat larger degree of linear polarization.

Recently we suggested using the temperature-polarization
cross correlation $\langle QT \rangle$ 
as a  further test ~\cite{cct94}.  Being linear 
in the polarization, this has some advantages for experiments
in which noise in the polarization is limiting, for the latter
averages to zero in $\langle QT \rangle$. 
The $\langle QT \rangle$ correlation also extends to larger angular scales 
than the auto-correlation function $\langle QQ \rangle$. 
 It thus has particular 
relevance to experiments with large sky coverage, such as 
post-COBE satellites now being planned.
Here we extend our calculations of $\langle QT \rangle$ to the tensor case,
with the result that a striking distinction emerges. 
On large angular scales the scalar 
and tensor $\langle QT \rangle$'s have the opposite
sign, as a
direct result of 
fundamental differences between
scalar and tensor modes.
Whether the rather small signal
we predict is observable is unclear,
depending primarily  upon the levels of
foreground contamination produced by our galaxy, which as
we shall discuss below is still unknown.

We begin with a simplified discussion.
To first approximation, 
recombination may be treated as instantaneous and 
there is no polarization.
The next approximation is to assume that each photon
subsequently undergoes a single scattering, after travelling
a comoving distance $\lambda$.
The photons we now receive from a given direction on the
sky emanated from a sphere of radius $\lambda$ surrounding the
scattering point (Fig. 1).
The Thomson cross section is 
$\sigma_T \propto |$$\boldmath{  \epsilon\cdot \epsilon'}$$|^2$, 
with $\boldmath {\epsilon}$
and $\boldmath {\epsilon'}$ the initial and final photon polarizations.
It follows that the 
photons we measure polarized along the $y$ axis mostly come from
above and below the scatterer, and
those 
 polarized along the $x$ axis
mostly come from the sides of the scatterer.
 
\begin{figure}[htbp]
\centerline{
\psfig{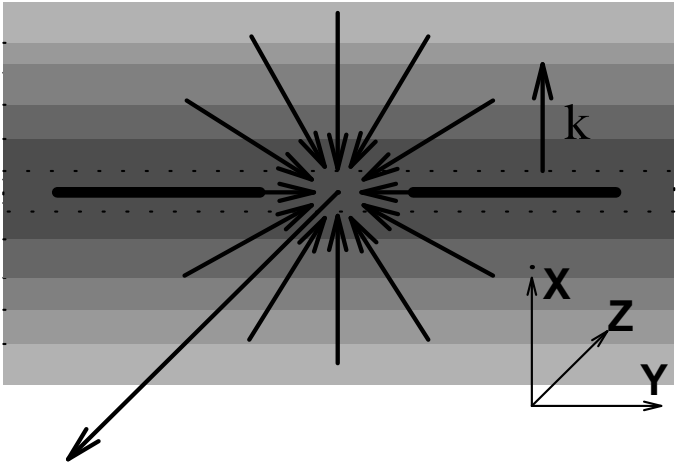}}
\vspace{-0.1 in}
\caption{A perturbation mode on the surface of last scattering,
with the paths of photons which we see coming from
a given point 
on the sky.
 For scalar perturbations, the direction
of linear polarization (heavy  lines) 
is aligned with the temperature troughs.
For tensor perturbations the polarization 
is instead perpendicular.  
}
\label{fig:fneil}
\end{figure}

In Figure  1
a single plane wave perturbation is shown, with its ${\bf k}$ vector 
perpendicular
to the line of sight. For
a scalar perturbation, the shading indicates the Newtonian 
potential $\Phi$, darker indicating 
more negative $\Phi$. 
Photons coming from the potential trough are
redshifted, so it appears to us as a temperature trough on the sky.
Photons coming from above and below the scatterer
fall into the potential trough  before they scatter, and so are 
relatively blueshifted. Photons from either side suffer
no such shift.
So the net linear polarization is aligned parallel to the temperature
troughs.
By rotating and superposing a set of such modes, one 
sees that temperature cold spots are surrounded by  a radial polarization 
pattern, hot spots by a tangential pattern.
A simple quantitative estimate may be made
in the long wavelength
limit ($ k\lambda \ll 1$), by averaging 
the Thomson cross section over the sphere of incident photons. 
The temperature anisotropy is
$\delta T/T  \approx {1\over 3} \Phi(s)$, with $s$ denoting
`at the scatterer'. But the polarization depends on the quadrupole moment of
$\Phi$ on the sphere - if we expand $\Phi(x,y)$ in a Taylor series, the
first terms that contribute involve second derivatives. 
One finds the fractional linear polarization measured using the
$x$, $y$ axes is
$Q=
-{1\over 20} \lambda^2(\partial_x^2 -\partial_y^2)(\delta T/T)(x,y)$,
where $T(x,y)$ is the two dimensional temperature field.

Gravitational waves are traceless, transverse excitations of the metric, 
$g_{\mu \nu}=a^2(\eta_{\mu \nu}+h_{\mu \nu})$, 
where $a$ is the cosmic scale factor.
Consider a 
 gravity wave with the same ${\bf k}$ vector and only 
$h_{zz}= -h_{yy}$ nonzero. 
The shading in Figure 1 indicates 
negative 
$h_{zz}$. 
The temperature distortion is given 
by the Sachs-Wolfe integral 
$-{1\over 2} \int h_{ij,0} n^i n^j \approx {1\over 2} h_{zz}(s)$,
with the dominant contribution
coming from the path shared by all photons from the scatterer to us.
So again Figure 1 shows a temperature trough on the sky.
But for gravity waves initially outside the horizon, 
if $h_{zz}$ is negative, $\dot{h}_{zz}$ is positive, so the proper distance is 
being expanded in the $z$ direction, contracted in the $y$ direction.
Now we see a fundamental difference with the scalar case.
Photons arriving at the scatterer from above and below are 
unaffected (since the gravity wave is transverse), whereas
photons coming from the two sides are blueshifted ($\dot{h}_{yy}$
being negative).  The induced polarization  pattern is  therefore
{\it perpendicular} to the temperature troughs.
 Again by rotating and 
superposing one sees that for gravity waves 
the temperature cold spots are surrounded 
by a tangential, hot spots by  a radial polarization pattern.  

Repeating the `single scattering' calculation explained above, 
we find
$Q \approx {1\over 20}  \lambda  h_{zz,0}(s).$
A
gravity wave mode  
initially outside the horizon evolves as 
$h_{zz} \propto 
1- A k^2 \eta^2$, for $k\eta \ll 1$, with $A={1\over 10}$ in the matter
era and ${1\over 6}$ in the radiation era. Using this, one finds
$Q  \approx {1\over 5} A
 (\partial_x^2-\partial_y^2) (\delta T/T)(x,y) \lambda \eta_{ls}$,
$\eta_{ls}$ being the conformal time at last scattering.

If the temperature
autocorrelation function is approximately scale invariant, 
then $\langle T (\theta, \phi) T 
(0) \rangle  \sim {\rm 
ln} (1/\theta)$, which is a rough 
approximation for both scalar and tensor cases 
for $\theta_{ls} \ll  \theta \ll 1$, $\theta_{ls}$
being the angle subtended by the horizon at last scattering.
Using the relation  between $Q$ and $T$ derived above, one sees
that $\langle Q (\theta, \phi) T (0) \rangle \sim
{\rm cos}(2 \phi) \theta^{-2}$, with coefficients of opposite sign for
the scalar and tensor cases. 
On smaller angular scales, one is sensitive to modes of
wavelength smaller than the horizon, which have begun to oscillate
by the time of last scattering. In the scalar case 
oscillations in the photon-baryon fluid density $\delta$ 
cause the sign of
$\langle QT \rangle$ to oscillate, since $T \propto \delta$ and $Q \propto 
\dot{\delta}$ are out of phase
~\cite{cct94}. Likewise in the tensor case, we have $T \propto h$ and
$Q \propto \dot{h}$, so the sign of $\langle QT \rangle$ also reverses.
However the redshifting away of gravity waves reduces these 
oscillations to negligible levels, so that $\langle QT \rangle$ is actually 
positive for
all $\theta \ll1$.  At very small angles, from 
the ${\rm cos} (2\phi)$ dependence and analyticity it follows that
$\langle QT \rangle$ vanishes like $\theta^2$.

To calculate $\langle QT\rangle$ accurately
we evolve the photon 
distribution function, ${\bf f}({\bf x},{\bf p},t)$, 
using the relativistic Boltzmann equation for radiative transfer with a 
Thomson source term, keeping terms to first order in the perturbation. 
The distribution function, ${\bf f}$, is a four dimensional vector 
describing the intensity and polarization degrees of freedom, with 
components related to the Stokes parameters: $T, Q, U,$ and $V$~\cite{ch}. 
The Boltzmann equation can be rewritten in terms of the perturbed 
distribution functions, or brightness functions, defined as, 
$\Delta^i \equiv 4 \delta f_i / \left( T_0 {{\partial \bar{f}}
\over {\partial T_0}} \right)$, where $T_0$ is the mean CBR temperature,
$\bar{f}$ is the unperturbed Planck distribution,
$\delta f_i$ is its first order perturbation, and
$i=T,Q,U$ or $V$. 

We evolve the coupled equations by expanding the perturbations in 
in plane waves.  
In the case of scalar perturbations, one needs to evolve only two 
transfer equations, 
those for the components of ${\bf f}$ corresponding to Stokes'
parameters $T$ and $Q$~\cite{k83}. 
In the tensor case the U component does not vanish but
Polnarev\cite{Polnarev} has shown  
with the proper choice of variables, 
\begin{eqnarray}
\Delta^T(\mu, \phi_{\bf k}) &\equiv& \alpha(\mu) (1 - \mu^2) \cos(2\phi_{\bf k}) \\
\Delta^Q(\mu, \phi_{\bf k}) &\equiv& \beta(\mu) (1 + \mu^2) \cos(2\phi_{\bf k}) \\
\Delta^U(\mu, \phi_{\bf k}) &\equiv& \beta(\mu) 2\mu \sin(2\phi_{\bf k})  
\end{eqnarray}
(where $\mu \equiv {\bf \hat{k}} \cdot{\bf \hat{q}}$ 
and $\phi_{\bf k}$ is the polar angle 
of ${\bf q}$ about ${\bf k}$) 
the four Boltzmann equations reduce to just two coupled equations:
\begin{eqnarray}
\dot{\alpha} + ik\mu\alpha &=& \dot{h} - \sigma_T n_e a[\alpha +\Psi]\\
\dot{\beta} + ik\mu\beta &=&  - \sigma_T n_e a[\beta -\Psi]
\end{eqnarray}
where, 
\begin{equation}
\Psi \equiv \frac{3}{32}\int d\mu'[(1 + 6\mu'^2 + \mu'^4) \beta(\mu') 
- (1-\mu'^2)^2\alpha(\mu')].
\end{equation} 
Here, 
$n_e$ is the density of 
free electrons. 
These equations are evolved, as in the scalar case\cite{be},
by expanding $\alpha$ 
and $\beta$ in Legendre polynomials
(i.e., $\alpha(\mu) = \sum_l (2l+1)\alpha_l P_l(\mu)$), 
converting them
into a hierarchy of ordinary differential equations \cite{critt93}.

Once these variables are evolved to the present epoch, correlation 
functions are evaluated by summing over ${\bf k}$ and possible polarizations.
One finds that the temperature-polarization cross correlation function is, 
\begin{eqnarray}
&&\langle Q({\bf \hat{q}}) T({\bf e_z}) \rangle = \frac{\cos 2\phi}{32 \pi^2} 
\int k^2dk \sum_{l,l'} (2l+1)(2l'+1) 
\nonumber \\ &&\hskip 20pt \times \biggl[ {(l'-2)!\over (l'+2)!}
[\alpha_{l'} B_l 
\cos^2\theta -A_{l'} B_l] a_{ll'} P^2_{l'}(\cos\theta) \nonumber 
\\~~ &&\hskip 40pt + \frac{1}{2}\sin^2\theta ~	
\alpha_{l'} B_l \bigl[\delta_{ll'}\frac{2}{2l+1} P_{l'}
(\cos\theta) \nonumber \\&&\hskip 100pt
+{(l'-4)!\over (l'+4)!}~  \tilde{a}_{ll'} P^4_{l'}(\cos\theta) \bigr]
\biggr],
\end{eqnarray}
where $(\theta$, $\phi)$ are the usual spherical polar angles of 
${\bf \hat{q}}$, and
the axes used to define the Stokes parameters are
${\bf e}_x$ and  ${\bf e}_y$.
Here, $A(\mu)\equiv\mu^2\alpha(\mu)$, $B(\mu) \equiv (1+\mu^2)\beta(\mu)$ and
the constants $a_{ll'}$ and $\tilde{a}_{ll'}$ are given by
$a_{ll'}=\int_{-1}^1dxP_{l}(x)P_{l'}^2(x)$,
 and
$\tilde{a}_{ll'}=\int_{-1}^1dxP_l(x)P_{l'}^4(x)$
which have simple closed form expressions. 

\begin{figure}[htbp]
\centerline{
\psfig{file=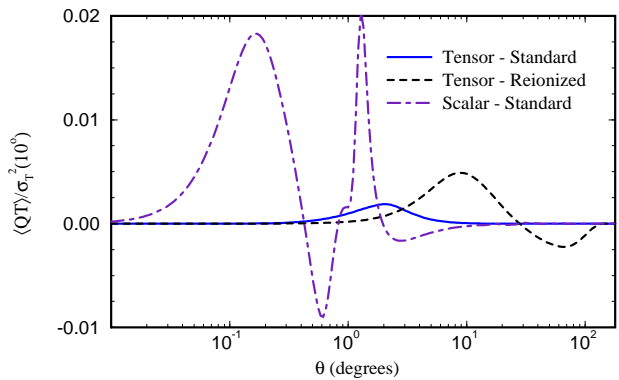,width=3.5in}}
\vspace{-0.75 in}
\caption{$QT$ correlation functions
for tensor perturbations in a universe with standard recombination, and
in a fully ionized universe, 
both
normalized assuming
that the entire COBE detection is due to gravity waves.
Shown for comparison is the result for scalar perturbations 
in a standard scenario~[11]. }
\label{fig:qtcorr}
\end{figure}
Figure~\ref{fig:qtcorr} shows $\langle QT \rangle$ for $\phi = 0.$ 
As for $\langle QQ \rangle$, 
on small scales the tensor $\langle QT \rangle$ is 
small in comparison to the scalar $\langle QT \rangle$.  At large 
$\theta$ , however, the signals have comparable magnitudes 
and opposite signs.  
With substantial early ionization, polarization is
greatly enhanced on large angular scales (note also that
a geometrical effect 
causes the tensor $\langle QT\rangle$ to reverse sign for $\theta > 30^o$.) 

\begin{figure}[htbp]
\centerline{
\psfig{file=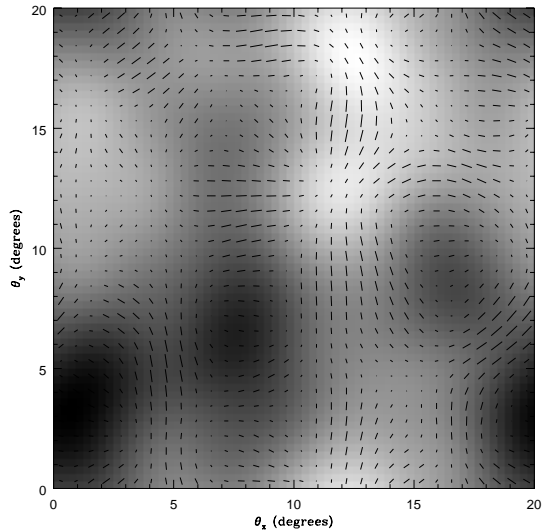,width=3in}}
\caption{$20^o \times 20^o$ temperature map resulting from 
tensor perturbations, smoothed on $3^o$, 
with the correlated component of the polarization
overlaid.}
\label{fig:tmap}
\end{figure}

\begin{figure}[htbp]
\centerline{
\psfig{file=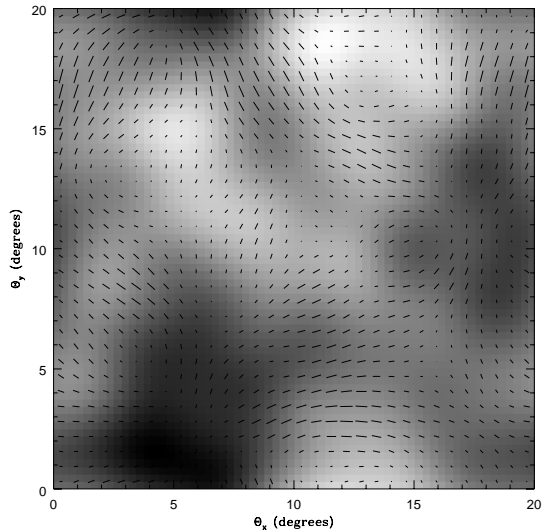,width=3in}}
\caption{As above, but for adiabatic  scalar perturbations.
}
\label{fig:smap}
\end{figure}
Since the primordial fluctuations are 
assumed to be described by gaussian statistics, the temperature and
polarization 
are completely described by 
$\langle QT \rangle$,
$\langle TT \rangle$ and
$\langle QQ \rangle$,
the latter two calculated in 
\cite{critt93} and \cite{cds} respectively. 
Using these it is straightforward to 
construct realizations of the microwave sky \cite{cct94}. 
The total polarization $Q$ is composed of parts which are correlated, 
$Q_C({\bf \hat{q}})$,  and 
uncorrelated, $Q_U({\bf \hat{q}})$,  with the temperature anisotropy. 
Figure~\ref{fig:tmap} 
shows the correlated component overlaid on  the temperature field. 
The length of each vector is proportional to
$[Q_C^2(\mbox{\boldmath$\theta$})+
U_C^2(\mbox{\boldmath$\theta$})]^{1\over2}$ and the orientation is given
by $2\phi=\tan^{-1}(U_C/Q_C)$.
Hot spots are seen to be associated with radial polarization patterns, 
while cold spots are surrounded by tangential polarization patterns. 
For scalar fluctuations on similar scales, however, the opposite is true
(Fig.~\ref{fig:smap}). 

\begin{figure}[htbp]
\centerline{
\psfig{file=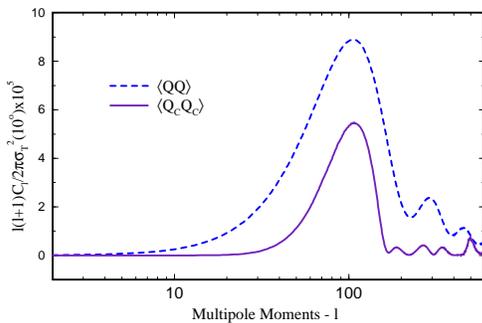,width=3in}}
\vspace{-0.75 in}
\caption{The power spectra 
 of the total (dashed line)
and temperature-correlated  (solid line) polarization
in a universe with a standard thermal history normalized to the COBE
measurement of the $10^{\circ}$ smoothed temperature variance.
}
\label{fig:qtpower}
\end{figure}

The power spectrum of $Q$ is compared to that for $Q_C$ in
Figure~\ref{fig:qtpower}, showing that in the tensor case 
the polarization is more strongly correlated with temperature 
than in the scalar case.
The variance $\sigma_{Q_C}^2$, proportional
to the area beneath the curve, comprises more than one third 
of the variance of the total polarization, $\sigma_{Q}^2$.
By comparison, for scalar fluctuations 
the correlated variance is barely a seventh of the total.

Are these temperature-polarization correlations observable? 
The main problem is whether sufficient
sky coverage can be obtained for a meaningful test.
To give ourselves the best possible chance we consider
an experiment which covers the whole sky, with very high 
signal to noise in the temperature measurement. 
A natural observable is then
 ${\cal O} = {1\over 4\pi} \int d\Omega Q Q_C$, being 
the integral of the `predicted' polarization $Q_C$ computed from the 
temperature alone, multiplied by the
`observed' polarization $Q$. 
 Since we are only
able to make a single measurement, a prediction for
${\cal O}$ is only testable if  
$S/N \equiv {\cal O}/ \sigma_{\cal O} \gg 1$.  
In Table~I  we have computed the signal to noise ratios for 
different beam sizes in a hypothetical full sky experiment.
These numbers include only the effects of cosmic variance -
instrument noise  may be modeled
by adding
an extra temperature-uncorrelated 
white-noise component to $Q$ with variance
$\sigma_N^2$. 
This has the effect of reducing 
$S/N$ by a factor $(1+\sigma_N^2/\sigma_Q^2)^{{1\over2}}$,
which may be computed from the values for $\sigma_Q$ given 
in the Table.

\begin{table}[htbp]
\begin{tabular}{|c|c|c|c|c|c|c|c|}
\multicolumn{2}{|c|}{Model} &
\multicolumn{3}{c|}{Standard} & 
\multicolumn{3}{c}{Reionized}  \\
\hline
& $\Theta $ &
$\sigma_Q$ &
$\sigma_{Q_C}$ &
$S/N$ &
$\sigma_Q$ &
$\sigma_{Q_C}$ &
$S/N$ 
\\
\hline
Scalar & 0$^{\circ}$ & 5.2 & 2.0 &  420
       & 3.1 & 1.1 &  26 \\
       & .5$^{\circ}$ & 1.3 & .53 & 91 
       & 3.0 & 1.0 &  24 \\
       & 1$^{\circ}$ & .59 & .26 & 49
       & 2.9 & .91 &  19 \\
       & 5$^{\circ}$ & .07 & .03 &  10 
       & 1.7 & .40 &  6 \\
\hline
Tensor & 0$^{\circ}$ & .46 & .28&  89 
       & 1.2 & .76 &  18 \\
       & .5$^{\circ}$ & .41 & .25 &  72 
       & 1.2 & .76 &  18 \\
       & 1$^{\circ}$  & .35 & .21 &  57 
       & 1.2 & .75 &  17 \\
       & 5$^{\circ}$ & .11 & .03 &  6 
       & .94 & .58 &  12 \\
\end{tabular}
\vskip 0.1in
\caption{Total and correlated polarization 
in different theoretical scenarios for various beam sizes $\Theta$ (FWHM).
$\sigma_Q$ and $\sigma_{Q_C}$ are given 
in $\mu K$, with the theories normalized to rms temperature 
fluctuations of $40 \mu K$ for  $\Theta= 10^o$. }
\end{table}

Apart from the forbidding challenge of building detectors with
the $\mu K$  sensitivities required,
the biggest question for using this technique to
 distinguish scalar and tensor perturbations
is whether the polarization signal is  
swamped by foreground contamination from
our galaxy. At frequencies below $\sim 100 $ GHz, synchrotron
emission is a significant background
to CMB experiments, and a typical expectation is
linear polarization at the level of
$5-10 \%$, significantly larger than the effects we are looking for 
(except perhaps in the
fully reionized scenarios).  At higher frequencies, dust emission becomes the
main background of concern. Here the situation appears much
more optimistic, 
with linear
polarization around $1\%$ being typical ~\cite{hildebrand93},
 although in 
special regions of very high magnetic field,
it can be substantially higher~\cite{morris92}.
The frequency dependence of the dust emission may allow 
one to subtract out dust-associated polarization using
multifrequency
measurements. In any case we hope the prospect of 
observing the intriguing signals
investigated here will stimulate 
further study of the likely backgrounds
 to
future full sky polarization measurements.

We thank R. Hills, A. Lasenby, M. Rees and D. Wilkinson for useful 
conversations.
RC thanks R. Davis and P. Steinhardt for collaborating in
the initial development of
the Boltzmann code.
The work of
DC was supported by a DOE 
grant (DOE-EY-76-C-02-3071) while that of 
RC and NT was partially supported by NSF contract
PHY90-21984, and the David and Lucile Packard Foundation.


\begin{thebibliography}{99}
\bibitem{smoot92}{G.F. Smoot {\it et. al.}, Ap. J. {\bf 396}, L1 (1992).}
\bibitem{critt93}R. Crittenden {\it et. al.}
Phys. Rev. Lett. {\bf 71}, 324 (1993).
\bibitem{infl}R.L. Davis {\it et. al.} Phys. Rev. Lett. {\bf 69} 1856 (1992);
L. M. Krauss and M. White, Phys. Rev. Lett. {\bf 69}, 863 (1992).  
\bibitem{edetal94}{E.J. Copeland {\it et. al.}, Phys. Rev. {\bf D48}, 2529
(1993).}
\bibitem{critt94} J.R. Bond {\it et.al.}, Phys. Rev. Lett.
{\bf 72}, 13 (1994); L. Knox and M. S. Turner, 
FERMILAB-PUB-94-175-A Preprint.
\bibitem{k83} N. Kaiser, Mon. Not. R. Astr. Soc. {\bf 202} 1169 (1983).
\bibitem{be} J.R. Bond and G. Efstathiou, Ap. J. {\bf 285}, L45 (1984);
Mon. Not. R. Astr. Soc. {\bf 226},
655 (1987).
\bibitem{Polnarev}{ A.G. Polnarev, Sov. Astr. {\bf 29}, 607 (1985);
R.A Frewin, A.G. Polnarev and P. Coles, MNRAS {\bf 266} L21 (1994).}
\bibitem{Ng93}K.L. Ng and K.W. Ng, Inst. Phys. Academica preprint 
IP-ASTP-08-93
\bibitem{cds} R.Crittenden, R. Davis and P.Steinhardt, Ap. J. {\bf 417},
L13 (1993).
\bibitem{cct94}D. Coulson, R.G. Crittenden and N.G. Turok, Phys. Rev. Lett.
{\bf 73}, 2390 (1994).
\bibitem{ch} S. Chandrasekhar, {\it Radiative Transfer} (Dover, New York, 1960)
pp. 1--53.
\bibitem{hildebrand93}{ R.H. Hildebrand {\it et. al.}
 Ap. J. {\bf 417}, 565 (1993).}
\bibitem{morris92} {M. Morris, Ap. J. {\bf 399}, L63 (1992).}
\end{thebibliography}
\end{document}